\documentclass[floatfix,twocolumn,preprintnumbers,amsmath,amsfonts,amssymb,notitlepage,showpacs,aps,prx,longbibliography,10pt]{revtex4-2}

\usepackage{xcolor}

\date{\today}

\usepackage{amsmath}
\usepackage{graphicx}
\usepackage{hyperref}
\usepackage{usebib}
\usepackage{bm}
\bibinput{bhm}

\usepackage[normalem]{ulem}

\begin{document}
\title{Multi-site gates for state preparation in quantum simulation of the Bose-Hubbard Model
}
\author{Pranjal Praneel}
\author{Thomas G. Kiely}
\author{Erich J Mueller}
\email{em256@cornell.edu}
\affiliation{Laboratory of Atomic and Solid State Physics, Cornell University, Ithaca, New York 14853}
\author{Andre G. Petukhov}
\email{apetukhov@google.com}
\affiliation{Google
Quantum AI, Santa Barbara, California 93111}

\begin{abstract}
We construct a sequence of multi-site gates which transform an easily constructed product state into an approximation to the superfluid ground state of the Bose-Hubbard model.  The mapping is exact in the one dimensional hard core limit, and for non-interacting particles in both one and two dimensions.   The gate sequence has other applications, such as being used as part of a many-body interferometer which probes the existence of doublons.
\end{abstract}
\maketitle

\section{introduction} 
State preparation is one of the most important challenges in analog quantum simulation.  The most common approach, adiabatic state preparation, performs poorly in gapless systems and cannot efficiently cross phase boundaries \cite{Dimitrova2023,Sorensen2010,RevModPhys.90.015002}.  Driven-dissipative approaches are promising, but require specialized hardware \cite{Harrington2022,Ma2019}.  Here we present a coherent approach to producing the ground state of the Bose-Hubbard model, based upon simple multi-site gates.  
These are more flexible than the two-site gates used in digital quantum simulation, and allow us to produce the exact ground state, without any Trotter error \cite{PRXQuantum.2.017003,Fauseweh2024,Smith2019,Heras2015}. { Related ideas have been proposed for producing states with  applications to quantum information theory \cite{Clark2005b,KAY2010,PhysRevA.90.042304}.}
Although our approach
can be used in a broad range of architectures \cite{PRXQuantum.2.017003}
it
is particularly appealing for simulation based on  transmon arrays, as all of the necessary  resources are found in the current generation of NISQ era quantum computers\cite{Roushan2017,PhysRevResearch.3.033043,PhysRevLett.126.180503,Yan2019,du2024probing,PhysRevLett.115.240501,li2024mapping,Gong2021,Kounalakis2018,roberts2023manybody,Saxberg2022}.

An analog quantum simulator uses tunable quantum hardware to mimic the behavior of a different system, or to realize a theoretical model of interest.  It can play a similar role to a wind-tunnel, allowing one to experimentally explore how a quantum model behaves as one changes parameters.  An example is using an array of coupled transmons to implement the Bose-Hubbard model \cite{fisher}:
\begin{equation}\label{bhm}
H=\sum_{\langle i j\rangle} g_{ij} (d_i^\dagger d_j+d_j^\dagger d_i) +\sum_i \delta_i d_i^\dagger d_i +\frac{\eta}{2} d_i^\dagger d_i^\dagger d_i d_i.
\end{equation}
Here $d_i,d_i^\dagger$ are ladder operators which change the number of excitations present in the transmon on site $i$, $g_{ij}$ parameterizes the strength of the coupling between sites $i$ and $j$, $\delta_i$ is the detuning, and $\eta$ encodes the transmon nonlinearity.  In modern hardware both $g_{ij}$ and $\delta_i$ are tunable, and can be made functions of time \cite{Kounalakis2018}.  If we interpret $d_i$ as the annihilation operator for a boson, Eq.~(\ref{bhm}) is the Bose-Hubbard model.  In that interpretation, $g_{ij}$ is a hopping matrix element, $\delta_i$ an on-site potential, and $\eta$ an on-site interaction -- more commonly written as $t_{ij}, V_i,$ and $U$.  The Bose-Hubbard model is one of the iconic models of strongly interacting quantum matter, displaying fascinating emergent properties, including a phase transition between a superfluid and insulating state\cite{fisher}.    While there are efficient classical methods to calculate any static properties of this model \cite{Pollet2013}, its dynamics are hard to calculate on a classical computer.  Thus this is a natural setting to demonstrate quantum advantage \cite{Preskill2018}.

Although the Bose-Hubbard Hamiltonian is natively realized by a transmon array, there is no well-established procedure for producing the superfluid ground state. {  One generic approach at unit filling is to turn off $g_{ij}$, and initialize the transmons in a product state, which can be chosen to be the ground state of Eq.~(\ref{bhm}) with $g_{ij}=0$}.  One then slowly ramps up $g_{ij}$.  By the adiabatic theorem, if this ramp is slow  compared to $\hbar/\Delta$, where $\Delta$ is the gap to the first excited state, then one should remain in the ground state.  Unfortunately in the thermodynamic limit this gap vanishes for the superfluid state.  Thus adiabatic state preparation is ill-suited for this task.

Our strategy will be similar, in that we will transform an easily produced product state into our state of interest.  Instead of adiabatic evolution, however, we will apply a set of multi-site gates, characterized by time evolution under Eq.~\ref{bhm} with carefully controlled $g_{ij}(t)$ and $\delta_i(t)$.  
In one dimension (Sec. \ref{sp}) our procedure produces the exact ground state for both very strong (hard core) and very weak interactions.  In two dimensions with repulsive interactions (Sec. \ref{gen2d}) it is only exact in the weakly interacting limit.  
A variant works in the strongly attractive limit (Sec. \ref{mps}), producing {\em cat} states where all particle sit on a single site, but the stack of particles is delocalized throughout the lattice.  
We also 
present further applications (Sec.~\ref{other}).

\begin{figure}
\includegraphics[width=0.8\columnwidth]{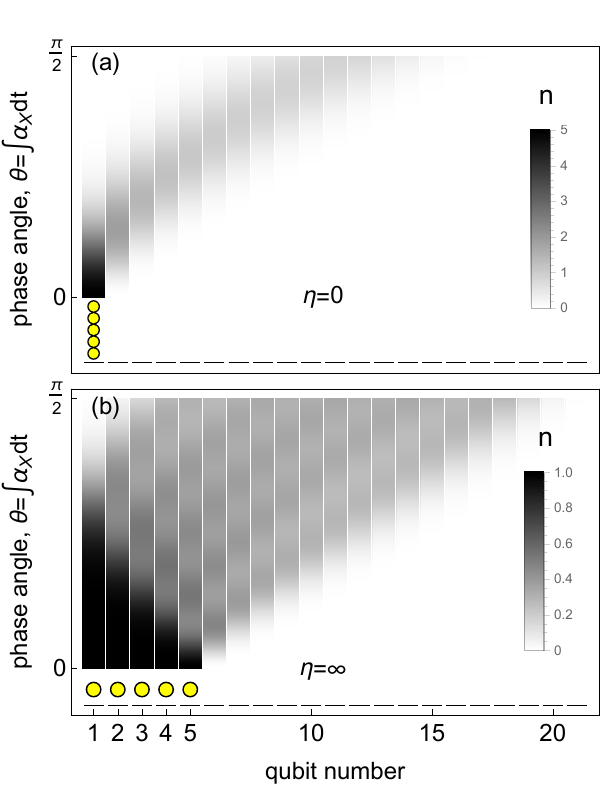}
\caption{Evolution of the boson density during the first pulse of our 2-pulse sequence in the
(a) non-interacting ($\eta=0$) and (b) strongly interacting ($\eta=\infty$) limit, illustrated here for $N=5$ particles on $L=21$ sites.  In (a) the initial state consists of all particles sitting on the left-most site. In (b), there is one particle per site on the 5 left-most sites.  After evolution by a phase angle of $\theta=\int \alpha_X(t) dt=\pi/2$, the density  matches that of the ground state of $H_X$ in Eq.~(\ref{hx}).  Our second pulse corrects the phases, producing the exact ground state. Note the different color scale for the two figures.}
\label{density}
\end{figure}

{
\section{One dimensional model}\label{sp}

\subsection{Single Particle Physics}\label{sp1d}
We begin by presenting a one dimensional model, where $g_{ij}$ vanishes unless $j=i+1$.  We will choose these matrix elements so that the hopping part of the Hamiltonian can be mapped onto a SU(2) spin matrix.  The structure of spin matrices allow us to construct a gate which produces the superfluid ground state of the Bose-Hubbard model.

We first consider the single particle limit, where there is only one excitation.  The Hilbert space is spanned by the states $|j\rangle=d_j^\dagger |{\rm vac}\rangle$, in which the particle is located on the site labeled $j$.  We will introduce a gate sequence which coherently transforms the single particle state $|j=1\rangle$, corresponding to a particle on the leftmost site of the chain, into a delocalized eigenstate of the hopping Hamiltonian
\begin{equation}
H=\sum_{\langle ij\rangle} g_{ij} (|i\rangle\langle j|+|j\rangle\langle i|).
\end{equation}
The many-body case will be discussed in Secs.~\ref{1d} where we include quantum statistics and interactions.

For a length $L$ chain, we introduce operators
\begin{eqnarray}\label{X}
X&=&\sum_{j=1}^{L-1}\sqrt{j(L-j)}\frac{|j+1\rangle\langle j|+|j\rangle\langle j+1|}{2}\\\label{Y}
Y&=&\sum_{j=1}^L\sqrt{j(L-j)}\frac{|j+1\rangle\langle j|-|j\rangle\langle j+1|}{2i}\\\label{Zop}
Z&=& \sum_{j=1}^L \left(j-\frac{L}{2}\right) |j\rangle\langle j|,
\end{eqnarray}
which are all of the form of Eq.~(\ref{bhm}), projected into the one-particle sector.  
These operators are constructed to form a spin $S=(L-1)/2$ dimensional representation of $SU(2)$.  Consequently
\begin{equation}\label{el1}
e^{-iZ\pi/2}e^{-iX\pi/2}Ze^{iX\pi/2}e^{iZ\pi/2}=X,
\end{equation}
corresponding to the fact that a $\pi/2$ rotation about the $\hat x$ direction, followed by a $\pi/2$ rotation about the $\hat z$ direction will map the $\hat z$ axis onto the $\hat x$ axis.  These gates are implemented by time evolving with appropriately selected $g_{ij}$'s and $\delta_i$'s.
They  map an eigenstate of $Z$ (which is easily produced) onto an eigenstate of $X$:
If $Z|\psi_0\rangle=E |\psi_0\rangle$ and $|\psi\rangle= e^{-iZ\pi/2} e^{-iX\pi/2}|\psi_0\rangle$, then $X|\psi\rangle=E|\psi\rangle$,
 which is the desired transformation.  
 
 { In particular, we could start with a particle on the left-most site of the lattice, which is the eigenstate of $Z$ with minimal eigenvalue.  It will be mapped onto to the eigenstate of $X$ with minimal eigenvalue.}

\subsection{Many-Body Physics of our 1D model}\label{1d}
In the many-body setting, the operators in Sec.~\ref{sp1d} generalize to
\begin{eqnarray}\label{cX}
{\cal X}&=&\sum_{j=1}^{L-1}\sqrt{j(L-j)}\frac{d_{j+1}^\dagger d_j +d_j d_{j+1}^\dagger}{2}\\\label{cY}
{\cal Y}&=&\sum_{j=1}^L\sqrt{j(L-j)}\frac{d_{j+1}^\dagger d_j-d_j^\dagger d_{j+1}}{2i}\\\label{Z}
{\cal Z}&=& \sum_{j=1}^L \left(j-\frac{L}{2}\right) d_j^\dagger d_j.
\end{eqnarray}
These obey the same algebra as $X,Y,Z$, forming a {\em reducible} representation of $SU(2)$.  The unitary operator $e^{-i{\cal Z}\pi/2} e^{-i{\cal X}\pi/2}$ will transform a many-body eigenstate of $\cal Z$ into a many-body eigenstate of $\cal X$.  The former corresponds to some set of localized particles, while the latter is analogous to momentum eigenstates.  

Including the role of interactions, 
we consider time evolution under the Hamiltonians
 \begin{eqnarray}\label{hx}
H_X&=&
\alpha_X(t)  {\cal X} + \frac{\eta}{2} H_{\rm int}\\\label{hz}
H_Z&=& 
\alpha_Z(t)
{\cal Z}+
\frac{\eta}{2} H_{\rm int}
\end{eqnarray}
where 
$H_{\rm int}=\sum_j d_j^\dagger d_j^\dagger d_j d_j$.
We then 
take $U_X$ 
and $U_Z$ 
to represent time evolution operators
under each of these Hamiltonians.  For example, $i\partial_t U_X=H_X U_X$.  Our goal will be to use $U_Z$ and $U_X$ gates to construct an approximation to the the ground state of $H_X$, starting from the ground state of $H_Z$.  In appendix~\ref{sec:uniform} we discuss the relationship to the uniform hopping model.

In the non-interacting limit, $\eta=0$, and the strongly interacting limit, $\eta\to\infty$, our approach will yield the exact many-body ground state of $H_X$.  In the absence of interactions $(\eta\to0)$ the Hamiltonians themselves obey the SU(2) spin algebra, and the arguments of Sec.~\ref{sp1d} apply directly.  When $\eta\to\infty$ we can instead use a Jordan-Wigner transformation to map the dynamics onto  non-interacting fermions.  The single-particle arguments from Sec.~\ref{sp1d} then apply in this fermionic representation.  Equivalently, one can note that the commutation relations of the operators $\cal X,Y,$ and $\cal Z$ are identical for bosons and fermions.

For the general case we need to introduce some extra notation.
Comparing Eq.~(\ref{hx}) with (\ref{bhm}), gives $g_{j,j+1}(t)=\alpha_X(t) \sqrt{j(L-j)}/2$.  { 
 The strongest coupling is { on the central bond,   $g_{\rm c}(t)=g_{L/2,L/2+1}(t)=\alpha_X(t)L/4$ for even $L$, and 
$g_{\rm c}(t)=g_{(L-1)/2,(L+1)/2}(t)=\alpha_X(t)
\sqrt{L^2-1}/4$ for odd $L$. } 
} 
Similarly for Eq.~(\ref{hz}) we have 
$\delta_j(t)=\alpha_Z(t) (j-L/2)$, and 
{ the largest detuning is found on the end qubits,
$\delta_L(t)=\alpha_Z(t) (L-1)/2$}.

We consider two easily prepared initial states, 
$|\psi_c\rangle$ 
and $|\psi_h\rangle$, 
both eigenstates of $H_Z$.  
The ``condensate wavefunction", $|\psi_c\rangle=(d_1^\dagger)^n|{\rm vac}\rangle/\sqrt{N!}$, has all $N$ particles on the left-most site of the lattice, and is the ground state of $H_Z$ for $\eta\to0$. The ``hard-core wavefunction", 
$|\psi_h\rangle=\prod_{j=1}^N d_j^\dagger|{\rm vac}\rangle$, has one particle per site on the left-most $N$ sites, and is the ground state of $H_Z$ for $\eta\to\infty$.  
 These initial states are illustrated in Fig.~\ref{density} (a) and (b).
Although the particles in %these initial states 
them are uncorrelated, the final states, which will be approximate eigenstates of $H_X$, are highly entangled.

\begin{figure}
\centering
\includegraphics[width=\columnwidth]{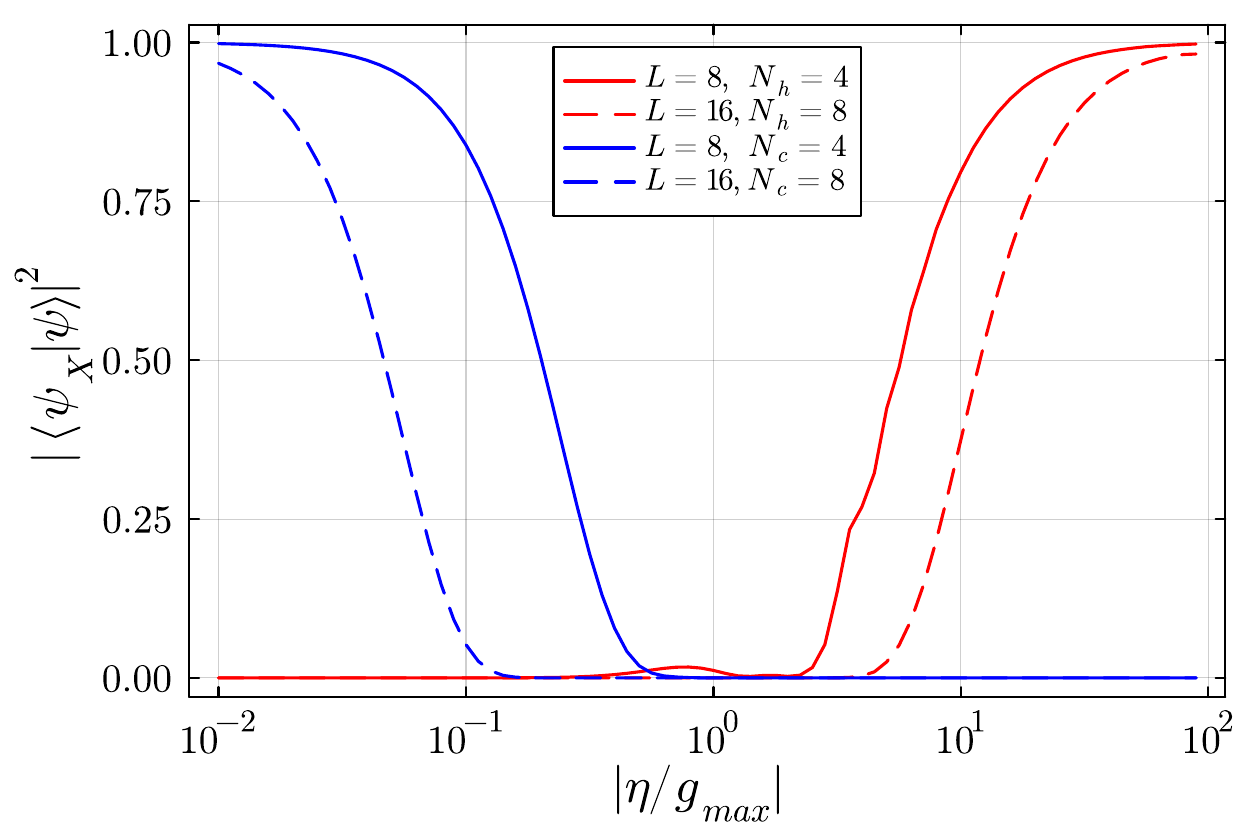}
\caption{
Fidelity $|\langle\psi_X|\psi\rangle|^2$ of state preparation using the pulse sequence in Sec.~\ref{1d} for the 1D Bose-Hubbard model.  Our target state, $\psi_X$, is the ground state of the Bose-Hubbard Hamiltonian $H_X$ in Eq.~(\ref{hx}, while $|\psi\rangle=U_Z(\varphi)U_X(\theta)|\psi_0\rangle$.  Two different initial states, $|\psi_0\rangle$, are considered, the condensate ansatz (blue, peaked on the left), $|\psi_c\rangle= (d_1^\dagger)^N|{\rm vac}\rangle/\sqrt{N!}$, and the hard-core ansatz (red, peaked on the right), $|\psi_h\rangle=\prod_{j=1}^N d_j^\dagger|{\rm vac}\rangle$.  Solid lines show $N=4$ particles in a chain of length $L=8$, while dashed lines have $N=8$ and $L=16$.  The pulse durations are selected so that $\theta=\varphi=\pi/2$.
}\label{FidelitySU2}
\end{figure}

We consider  pulses 
of the form
\begin{align}\label{gm}
\alpha_X(t)&=\alpha_X^{\max }f(t,\tau_\theta)\\
\alpha_Z(t)&=
\alpha_Z^{\rm max}
f(t,\tau_\varphi),
\end{align}
where $f(t,\tau)$ is a Gaussian-broadened
 box-shaped
waveform, which is approximately 1 over an interval of length $\tau$, and is otherwise zero,
\begin{equation}
f(t,\tau )=\frac{1}{2} \left(\text{erf}\left(\frac{t-p \sigma }{\sqrt{2}
   \sigma }\right)-\text{erf}\left(\frac{t-p
   \sigma -\tau}{\sqrt{2} \sigma
   }\right)\right).    
\end{equation}
Here $\sigma$ is the rise and fall times, and $p$ is a padding parameter chosen so that $f$ effectively vanishes for negative $t$.  This pulse shape interpolates between a square wave and a Gaussian as $\sigma$ is varied.
It is normalized to satisfy $\tau^{-1}\int_0^T f(t,\tau)dt=1$, where $T$ is the pulse duration, including the rise and fall times.  In our numerics we take $T=\tau+2p\sigma$ and $p=5$.  { This value of $p$ is chosen so that the fractional phase error from truncating the pulse is less than $10^{-6}$.}  We take $\sigma=\tau/10$, but verified that the results are only weakly dependent on the rise/fall time.  We include it, however, to more accurately model the pulses used in experiments.  We define $g_{\rm max}$ and $\delta_{\rm max}$ to be the largest value of the coupling or detuning during the sweep.

We define phase angles
\begin{align}
\theta &= \int\! \alpha_X(t)\,dt=\frac{4 \tau g_{\rm max}}{L},\frac{4 \tau g_{\rm max}}{\sqrt{L^2-1}}\\
\varphi &= \int\! \alpha_Z(t)\,dt=2 \tau \delta_{\rm max}/(L-1),\label{vf}
\end{align}
where the two expressions for $\theta$ correspond to even or odd particle numbers.  

Figure~\ref{density} shows how the density evolves during the $U_X$ operation, in both the non-interacting (a) and hard-core (b) limit.  In both these cases the evolution only depends on $\theta$, and not  the detailed pulse shape.  The particle distribution evolves from highly localized to symmetrically distributed.  The $U_Z$ operation does not change the density profile, but simply introduces phases in the wavefunction's coherences.

We quantify the accuracy of the gate by the fidelity $|\langle\psi_X|\psi\rangle|^2$, where $|\psi_X\rangle$ is our target state.  It is the lowest eigenstate of $H_X$. The state produced by the pulse sequence is {
\begin{equation}
|\psi\rangle=U_Z U_X |\psi_0\rangle.
\end{equation}}
Figure~\ref{FidelitySU2} shows the fidelity as a function of $\eta/g_{\rm max}$, using $\delta_{\rm max}=g_{\rm max}$ and $\theta=\varphi=\pi/2$.  Exploration of other phase angles can be found in Appendix~\ref{optimize}. %and $\delta_{\rm max}=g_{\rm max}$.
Typical experiments have $g/(2\pi)\sim 10$MHz, yielding a gate time of $T \sim 0.1 \mu$s.  

As expected, in both the non-interacting limit $\eta/g\to 0$ and the hard-core limit $\eta/g\to\infty$ the fidelity approaches unity, assuming we choose the appropriate starting state.  Importantly, one can produce the ground state with very high fidelity over a large range of $\eta/g$.  In typical transmon experiments the nonlinearity dominates over the coupling, $\eta/g\gtrsim30$, and the fidelity will be extremely high.    We give details about our numerical techniques in Appendix~\ref{numerics}.

{An important feature in Fig.~\ref{FidelitySU2} is the size dependence of the  fidelity.  This can be quantified by finding the value of $|\eta/g_{\rm max}|$ where $|\langle\psi_X|\psi\rangle|^2=0.5$.  We find that this cross-over scale appears to scale as $|\eta/g_{\rm max}|\sim L$ for the hard-core initial conditions and $|\eta/g_{\rm max}|\sim L^{-2}$ for the condensate initial conditions.  }

\begin{figure}
    \centering
\includegraphics[width=\linewidth]{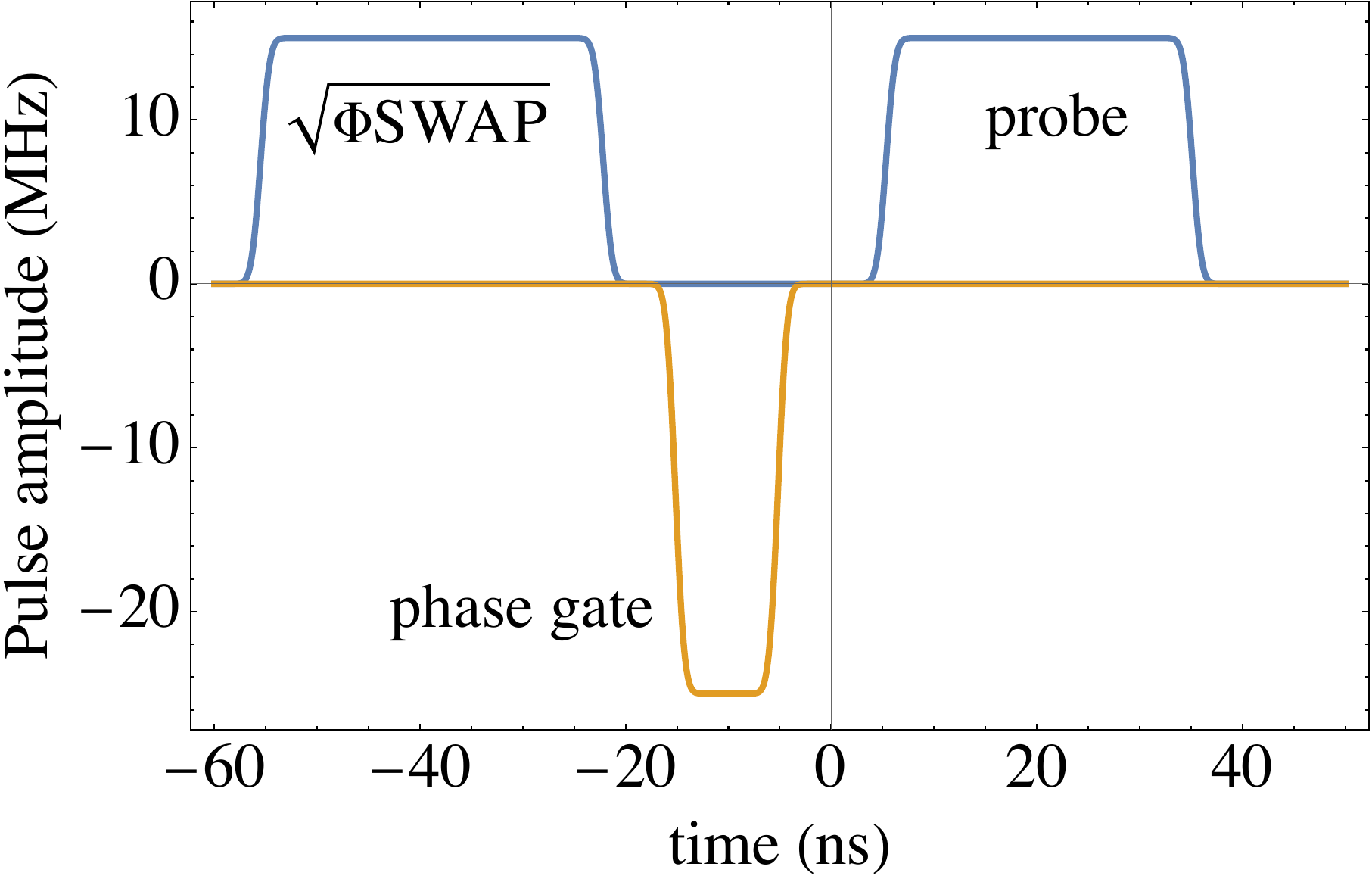}
    \caption{Pulse sequence for preparing and probing a one-dimensional chain with $L=8$ and $N=4$ by applying the set of multi-site gates
    described in the text.  { The blue curves, plotted on the positive axis, represent the coupling on the central bond, $g_{\rm c}(t)$, whose maximum value is $g_{\rm max}$.  The yellow curve, plotted on the negative axis, corresponds to the  detuning of the edge qubit, $\delta_{L}(t)$, whose maximum value is $\delta_{\rm max}$.}}
    \label{Fig:Sequence}
\end{figure}

To experimentally verify that one has produced the desired state, we propose looking at the time evolution of the density profile under $H_X$.  Figure~\ref{Fig:Sequence} shows the required pulse sequence, starting from the product state $|\psi_0\rangle$.  In this schematic, the $U_X$ operation is labeled $\sqrt{\Phi SWAP}$, the $U_Z$ operation is labeled ``phase gate", and ``probe" labels time evolution under $H_X$.  (See Sec.~\ref{other} for more discussion of this notation.)  The vertical axis shows the magnitude of $g_{\rm c}(t)$ and $\delta_{L}(t)$.  The first two pulses correspond to our gate sequence, while the third is used to verify that we have created the desired state.  In particular, during this probe pulse the density profile will be static if and only if the system is in an eigenstate of $H_X$.

To quantify the density evolution during the probe pulse, we introduce the center-of-mass operator $\zeta=(8/L^2)Z$.  If the system is prepared using an imperfect set of gates, e.g.
the angle of the phase gate $\varphi$ is not exactly $\pi/2$, the ``magnetization'' vector $\langle{\bm M}\rangle=(\langle X \rangle,\langle Y  \rangle, \langle Z \rangle)$ will undergo a Larmor precession around $x$-axis.  In the hard-core limit one expects  that $\langle \zeta \rangle=\cos(\varphi)\sin(\alpha_X t_h)$,
where $\alpha_X$ is the coefficient multiplying the hopping term in Eq.~(\ref{hx}), and $t_h$ is the duration of the probe pulse.  Similar results will be found outside of the hard core limit:  If one is not in an eigenstate of $H_X$,
the measured value of $\langle z \rangle$ will 
reveal pronounced Rabi flops.  This structure is shown in Fig.~\ref{Fig:COM} for experimentally relevant parameters.
The absence of the Rabi oscillations in the dependence of $\langle \zeta \rangle$ on $t_h$ will be indicative of the high quality  state preparation.  {  For comparison, the fidelity is a quadratic function of $\varphi$, peaked at $\varphi=0$.}

As discussed in appendix~\ref{optimize}, for finite $\eta/g$ the optimal state preparation gates require $\varphi=\varphi^*<\pi/2$.  As seen in Fig.~\ref{Fig:COM}, the Rabi oscillations { nearly} disappear when the gate angle is tuned to this optimal value.
}

\begin{figure}
    \centering
\includegraphics[width=\linewidth]{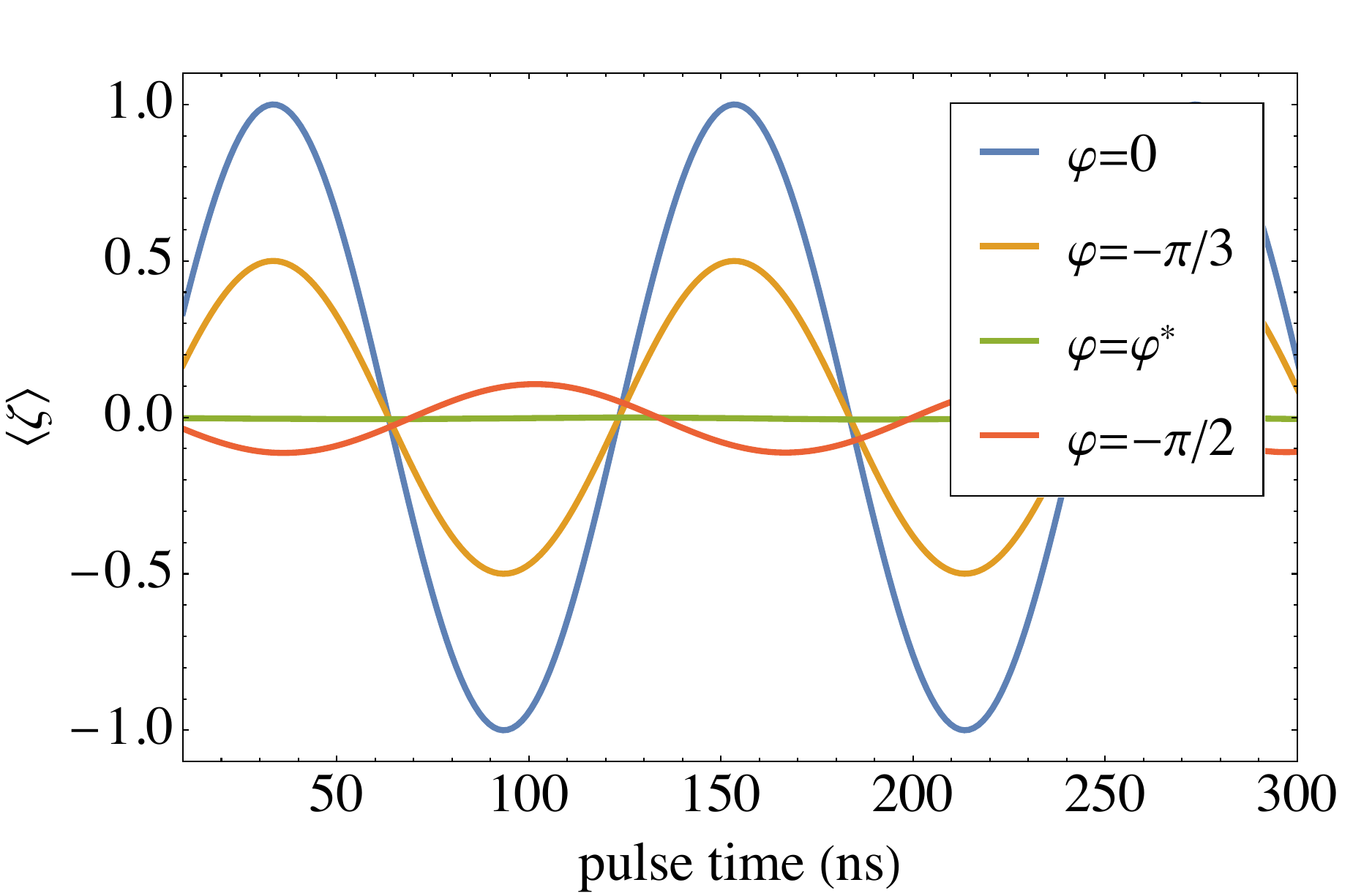}
    \caption{{ Expectation value of the center-of-mass operator $\zeta$ obtained numerically for the Bose-Hubbard Hamiltonian of a linear chain with $L=8$ and
    $N=4$, as a function of the duration of the probe pulse in Fig.~\ref{Fig:Sequence}.  Here  $\eta/2\pi=-270$~MHz{, $g_{\rm max}/2\pi=15$MHz,
    $\delta_{L}/2\pi=25$MHz
    }. The numerical results are in a good agreement with expected sinusoidal dependence 
    predicted in the hardcore limit. Similarly to this prediction, we also see an {\em almost} complete suppression of the Rabi oscillations indicative of the high fidelity state preparation. As further explored in Appendix~\ref{optimize}, this suppression occurs at the optimal $\varphi^*=-0.47\pi$ rather than at the nominal $\varphi=-\pi/2$ expected in the hardcore limit.  {This optimal angle depends on the ratio $\eta/g_{\rm max}$.}}}
    \label{Fig:COM}
\end{figure}

{

}

\section{Two Dimensions}\label{gen2d}
\subsection{Single Particle Physics}\label{sp2d}
We now generalize the arguments from Sec.~\ref{sp} to two dimensions.  As before, we begin by considering the single particle limit.
The construction in Sec.~\ref{sp1d} can be extended to two dimensions by considering the SU(3) algebra instead of SU(2).  The argument is most transparent in the 
Schwinger boson construction of the $d(\ell,0)$ representation \cite{Fradkin1965}.  { Related constructions were discussed in \cite{Post2014,PhysRevA.85.062306,Genest2013,Genest2015}.}

\begin{figure}
\includegraphics[width=0.8\columnwidth]{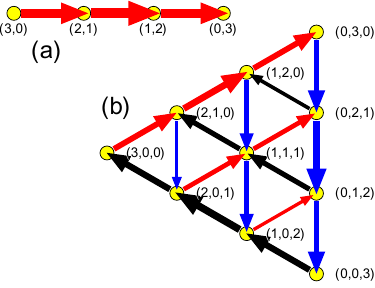}
\caption{
(a) One dimensional lattice labeled by non-negative integers $(n_a,n_b)$ with $n_a+n_b=\ell$. (b) Two dimensional lattice labeled by non-negative integers $(n_a,n_b,n_c)$ with $n_a+n_b+n_c=\ell$.
 In this case $\ell=3$.  Illustrated are the hoppings $b^\dagger a, c^\dagger b, a^\dagger c$ as red, blue, and black arrows.  The arrow width is proportional to the matrix element.  For example, when you move along a red arrow, $n_a$ is reduced by 1 and $n_b$ is increased by 1, and the hopping amplitude is $\sqrt{n_a(n_b+1)}/2$ in 1D or $\sqrt{n_a(n_b+1)}/3$ in 2D. }
\label{triangle}
\end{figure}

As preparation, we first reparameterize the model from Sec.~\ref{sp1d} in terms of Schwinger bosons \cite{Schwinger1952}.
 Rather than labeling the {\em single particle states} with a single integer, one uses
pairs of non-negative integers $|n_a,n_b\rangle$ with the constraint $n_a+n_b=\ell=L-1$.  We identify $|j\rangle=|n_a,n_b\rangle$ where $n_a=j-1,$ and $n_b=\ell-j+1$. For example, if we have 4 sites, we would label them as $|0,3\rangle,|1,2\rangle,|2,1\rangle,|3,0\rangle$ -- see Fig.~\ref{triangle} (a).

This rewriting simplifies the book-keeping, allowing us to introduce ladder operators $a$ and $b$: $a|n_a,n_b\rangle=\sqrt{n_a}|n_{a}-1,n_b\rangle$ and $b|n_a,n_b\rangle=\sqrt{n_b}|n_{a},n_b-1\rangle$. 
 These operators take us out of the physical Hilbert space, as the sum $n_a+n_b$ is reduced by 1.
Bilinears $a^\dagger a$, $a^\dagger b$, $b^\dagger a$ and $b^\dagger b$, however, keep us in the physical space.

We then note that the operators in Eqs.~(\ref{X}) through (\ref{Zop}) can be expressed as
\begin{eqnarray}
X&=&(b^\dagger a+a^\dagger b)/2\\
Y&=&(b^\dagger a-a^\dagger b)/2i\\
Z&=&(b^\dagger b-a^\dagger a)/2.
\end{eqnarray}
  The mapping in Eq.~(\ref{el1}) is then readily derived by using the algebra of raising and lowering operators (see Appendix~\ref{su2}). 
  The bilinears $S_-=a^\dagger b$ and $S_+=b^\dagger a$ act as hopping operators on the 1D chain, that respectively move the particle to the right or left.

One possibly confusing feature of this notation is that the ladder operators $a$ and $b$ do not change the number of particles in the system -- rather they change the integers $n_a$ and $n_b$ which label the lattice sites.  It is the operators $d_j^\dagger$ and $d_j$ that add or remove  particles.  As was clear from the discussion in Sec.~\ref{1d}, implementing the operators in Eq.~(\ref{X}) through (\ref{Zop}) requires spatially dependent nearest-neighbor couplings $g_{ij}$ or biases $\delta_i$.

We generalize this argument to 2D by considering 
a triangular array of sites, labeled by three non-negative integers, $|n_a,n_b,n_c\rangle$ with the constraint that $n_a+n_b+n_c=\ell$ -- see Fig.~\ref{triangle} (b).  We again introduce ladder operators $a,b,c$.  As before, these operators do not change the number of particles, but instead change the integers $n_a,n_b,n_c$ which label the single particle states.  

As with the SU(2) case, physical operators can be constructed from the product of one creation operator and one annihilation operator.
 In particular the six bilinears, $a^\dagger b,b^\dagger c, c^\dagger a, b^\dagger a, c^\dagger b, a^\dagger c$, are hopping operators, moving a particle around the lattice.  
The combinations 
\begin{eqnarray}
Q&=&a^\dagger a-{\ell/3}\\ 
W&=&\frac{1}{3}(a^\dagger+b^\dagger+c^\dagger)(a+b+c)
{-\frac{\ell}{3}}
\end{eqnarray}
play the role of that $Z$ and $X$ performed in Sec.~\ref{sp1d}:  $Q$ corresponds to a lattice tilt, while $W$ represents a hopping Hamiltonian. 
One may also express $Q$ and $W$ in terms of the standard Cartan-Weyl basis of SU(3) \cite{Cahn1984-rt}.  Up to constants, $Q$ is the hypercharge, $Y$, while $W$ is the sum of the raising and lowering operators $T_{\pm},U_{\pm},V_{\pm}$.  

Similar to the case in Sec.~\ref{sp1d}, the hopping matrix elements from $W$ are inhomogeneous.   For example, $b^\dagger a |n_a,n_b-1,n_c\rangle=\sqrt{n_b n_a}|n_a-1,n_b,n_c\rangle$.  As depicted in Fig.~\ref{triangle}, the hopping is strongest along the perimeter.

The eigenvalues of $Q$ are just the integers $n_a=0,1,2,\cdots \ell$.  They have degeneracies $\ell+1-n_a$, corresponding to the number of ways of choosing $n_b$ and $n_c$ with the constraint that $n_a+n_b+n_c=1$.  Given that $W$ is related to $Q$ by a unitary rotation, they have identical spectra.

Straightforward algebra (see Appendix~\ref{su3}) gives
\begin{eqnarray}
e^{-iQ 2\pi/3}e^{-iW 2\pi/3}Qe^{iW 2\pi/3} e^{iQ 2\pi/3}&=&W,
\end{eqnarray}
yielding a set of gates will convert an eigenstate of $Q$ into an eigenstate of $W$.  
Thus if we started with a single particle on a site with quantum number $n_a$, it would evolve to a delocalized state, which is an eigenstate of the hopping Hamiltonian  $W$.

\subsection{Many-Body Physics of our 2D Model}\label{2d}
We repeat the analysis of Sec.~\ref{1d}, but with the 2D model in Sec.~\ref{sp2d}.  We define
\begin{eqnarray}
{\cal Q}&=&\!\!\!\!\!\!
\sum_{n_a+n_b+n_c=\ell}\!\!\!
(n_a\!-\!{\ell/3})
d^\dagger_{n_a,n_b,n_c} d_{n_a,n_b,n_c}\\ 
{\cal W}&=&\frac{1}{3}
\!\!\!\!\!\!\!\!\!\sum_{n_a+n_b+n_c=\ell}\!\!\!\!\!\!\!\!\!\sqrt{n_a n_b}\,\, d^\dagger_{n_a-1,n_b+1,n_c}d_{n_a,n_b,n_c}+{\rm sym}\nonumber
%+\frac{\ell N}{3}
\\
\end{eqnarray}
where $d_{n_a,n_b,n_c}$ is the annihilation operator for the site labeled by integers $(n_a,n_b,n_c)$ with $n_a+n_b+n_c=\ell$.  In the definition of $\cal W$,
only the first of the 6 symmetry related hopping terms are shown,  and $N$ is the total number of particles. 
We construct the Hamiltonians
\begin{align}
H_W &= 
\alpha_W(t)
{\cal W}-\frac{\eta}{2}H_{\rm int}\\
H_Q &= 
\alpha_Q(t)
{\cal Q}-\frac{\eta}{2}H_{\rm int}
\end{align}
and again define the largest spatial coupling to be $g_{\rm c}(t)=\alpha_W(t) (\ell+1)/6$,  or $g_{\rm c}(t)=\alpha_W(t) \sqrt{\ell( \ell+2)}/6$ for odd or even $\ell$. The largest detuning is $\delta_{L}(t)=\alpha_Q(t) {\times(2\ell/3)}$, and $\theta$ and $\varphi$ by Eqs.~(\ref{gm})-(\ref{vf}) -- replacing $X$ by $W$ and $Z$ by $Q$.  In this case, however, we take $\theta=\varphi=2\pi/3$, corresponding to the gates in Sec.~\ref{sp2d}.  During a pulse, the  largest value of $g_{\rm c}(t)$ and $\delta_{L}(t)$ will be $g_{\rm max}$ and $\delta_{\rm max}$.
We again consider two initial states which are eigenstates of $H_Q$:  Consisting of either all $N$ particles on the same site ($\psi_c$), or divided between the $N$ left-most sites ($\psi_h$).

\begin{figure}
\centering
 \includegraphics[width=
 \columnwidth]{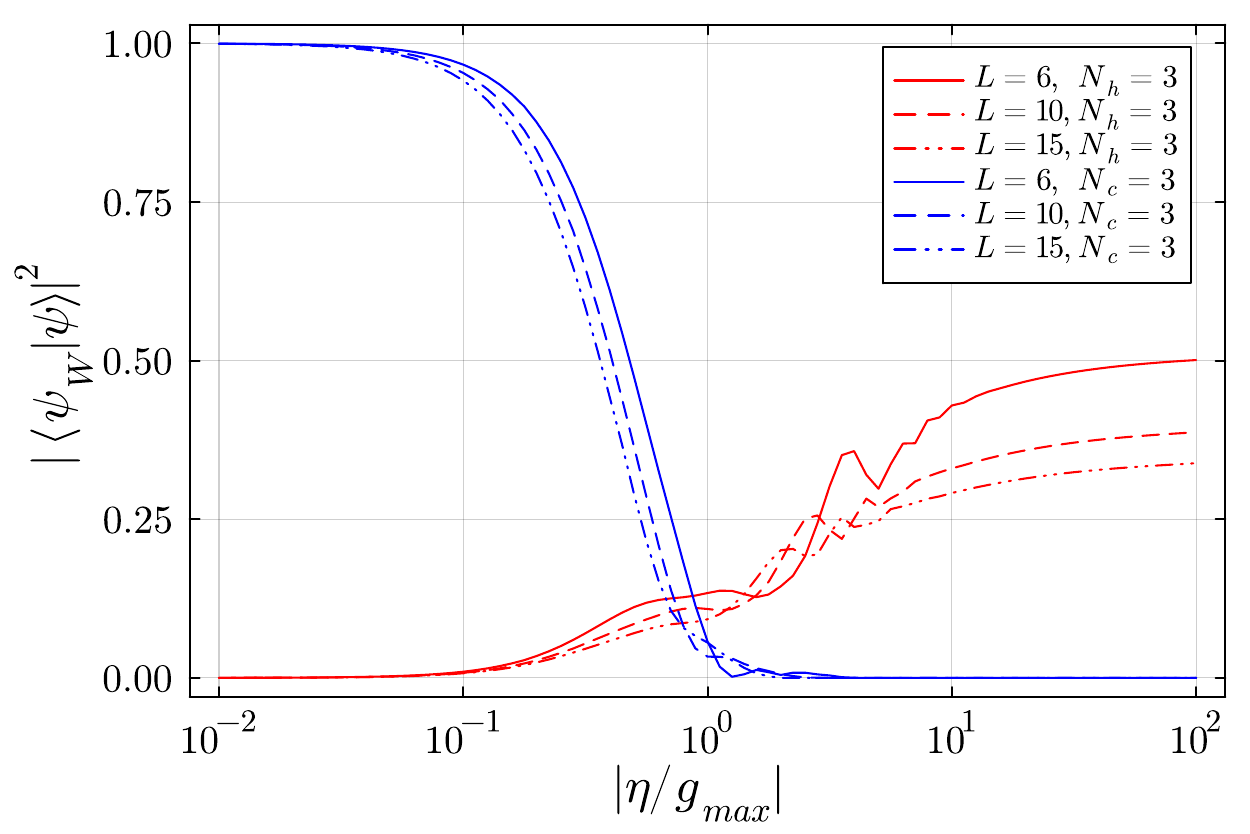}
\caption{Fidelity of state preparation on the 2D triangular lattice with $g<0$ and $\eta>0$.  Here $\ell=2,3,4$, corresponding to $L=6,10,15$ sites with  $N=3$ particles. Initial state is (left, blue) $|\psi_c\rangle$ or (right, red) $|\psi_h\rangle$.%{This calculation is valid only if $\eta/g <0$! Otherwise we cannot start with a product state. The ground state is degenerate at $g=0$ and any infinitesimal $g$ leads to highly entangled eigenstates. Discuss!}
}
\label{mboverlap}
\end{figure}

Figure~\ref{mboverlap} shows the state-creation fidelity for a lattice with { $\ell=2,3,4$, corresponding to $L=(\ell+1)(\ell+2)/2=6,10,15$ sites.  We take $N=3$, $g<0$, and $\eta>0$.  Again we find that starting from $|\psi_c\rangle$ we produce the exact ground state of $H_W$ as $\eta/g\to0$.  The large fidelity in this limit is expected, as each particle is independent.

In the hard core limit $\eta/g\to\infty$, starting with $|\psi_h\rangle$, the fidelity approaches 0.51,0.39,0.34
%0.5094,0.3924,0.3412 
for $L=6,10,15$.  This is not as high as the 1D case, as there is no exact mapping of 2D hard core bosons onto non-interacting fermions. 
{  For fixed $L$ the fidelity is a non-monotonic function of $N$.  In the hard-core limit it is largest when either the density of particles or holes is small.  We would argue that} this two dimensional protocol is best reserved for platforms which realize the weakly interacting system.

\section{Multi-particle bound states}\label{mps}

In the strongly interacting limit $|\eta|\gg |g|,|\delta|$ we can also consider multi-particle bound states where {\em all} of the particles sit on the same site -- but that bound complex can be on a superposition of sites.  These {\em cat} states are highly excited eigenstates of the {\em repulsive} Bose-Hubbard model, and form the true ground state of the {\em attractive} model \cite{PhysRevB.103.L220202,PRXQuantum.3.040314}. In what follows we will consider the latter, i.e. use Hamiltonian~\eqref{bhm} with $\eta <0$ and $g_{ij}<0$.

We define $|j\rangle$ to be the state where all particles sit on site $j$.  Integrating out the configurations where fewer than $N$ particle are on a single site, we { use $N$'th order perturbation theory to}  find an effective model Hamiltonian 
describing dynamics within the manifold of the cat states:
\begin{equation}\label{heff}
H_{\rm eff}\!=\!\sum_{ij} g_{ij}^{\rm eff}(t) |i\rangle \langle j| +h.c.+ \sum_iV_i(t)| i\rangle\langle i |
 \end{equation}
where
\begin{eqnarray}\label{gijeff}
g_{ij}^{\rm eff}(t)&=&-\frac{N }{(N-1)!|\eta|^{N-1}}\!|g_{ij}^N(t)| \\\label{Eq:Vi}
V_i(t)&=&N\delta_i(t)-\frac{N }{(N-1)|\eta|}\sum_j g_{ij}^2(t)+\cdots
\end{eqnarray}
with $g_{ij}$ and $\delta_i$ corresponding to the Hamiltonian terms in Eq.~(\ref{bhm}).  
The correction terms in $V_i$ come from virtual processes where the complex breaks up, and then reforms on its original site.  {  See \cite{PhysRevB.103.L220202} for a full derivation}.

By appropriately choosing coupling factors $g_{ij}$ and on-site biases $\delta_i$ we can make $H_{\rm eff}$ proportional to the operators $X,Z,W,Q$. We can then produce a gate sequence which, in the case of strong attraction, converts a localized state into the cat ground state of $X$ or $W$.  

In all cases we take pulses where the pattern of weights is fixed, but the amplitudes vary, $g_{ij}(t)=g_{\rm c}(t) w_{ij}$, 
and $g^{\rm eff}_{ij}(t)=
g_{\rm c}^{\rm eff}(t) 
w_{ij}^{\rm eff}$, 
where the largest $w_{ij}, w_{ij}^{\rm eff}$ is unity.  For example, in the 1D case with even $L$, the $X$ gate requires
$w^{\rm eff}_{j,j+1}=2\sqrt{j(L-j)}/L$ and hence we choose
 $w_{j,j+1}=(2\sqrt{j(L-j)}/L)^{1/N}$.  We use the pulse shapes described in Sec.~(\ref{1d}), with $g_{\rm c}(t)=g_{\rm max} f(t,\tau_\theta)$,  first applying a pulse for which $H_{\rm eff}$ is proportional to $X$, then where $H_{\rm eff}$ is proportional to $Z$.  
 
 In 2D we follow the same procedure, but with an effective $W$ gate followed by an effective $Q$ gate.  For the effective $W$ gate on the triangular lattice, the weight factor for the bond with $i=\{n_a,n_b,n_c\}$ and
 $j=\{n_a-1,n_b+1,n_c\}$  is
 $w_{ij}=\left(\xi(\ell)\sqrt{n_a(n_b+1)}\right)^{1/N},$
where the normalization factor is $\xi(\ell)=6/(\ell+1)$ or $6/\sqrt{\ell(\ell+2)}$ for odd or even $\ell$.

In Fig.~\ref{FidelityAttraction} we show the overlap between the ground state of Eq.~(\ref{bhm}) with attractive interactions, and the state produced on a triangular when we apply our pulses with optimize phases $\theta$ and $\phi$ (cf. Fig.~\ref{bhc}).  As can be seen, we we find excellent fidelities as long as $g_{\rm max}<0.05\times \eta$.

\begin{figure}
\centering
\includegraphics[width=\columnwidth]{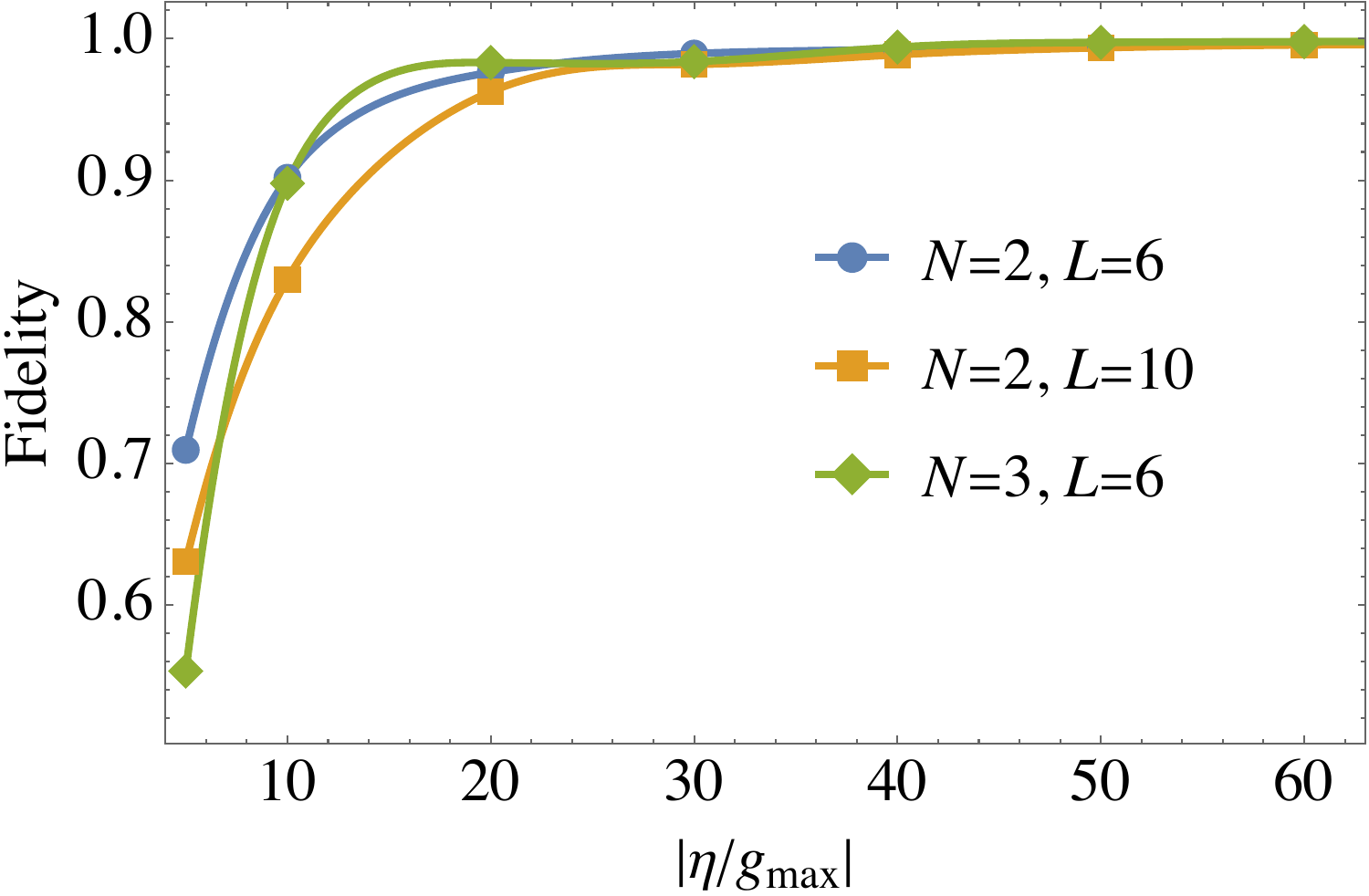}
\caption{Fidelity of the groundstate preparation of  the attractive 2D Bose-Hubbard model using the protocol in Sec.~\ref{mps}.  Here  we included terms up to second order in $g$ in Eq.~(\ref{Eq:Vi}) and optimized the phase angles $\varphi,\theta$ to achieve a high
quality ground state.
}\label{FidelityAttraction}
\end{figure}
One challenge is that the pulse times are set by the inverse of $g_{\rm eff}^{\rm max}$, and hence 
scale as $\eta^{N-1}/(g^{\rm max})^N$.  Thus these times 
become very long when $g^{\rm max}$ is small and $N$ is large.
In practice, however, the times are reasonable for $N\sim 3$.   
If we take $g_{\rm max}/(2\pi)=25$~MHz, $\eta/(2\pi)=-300$~MHz, $N=3$, and $\ell=2$, we find $\tau=522$~ns for an effective $W$ gate.
To explore this physics 
for larger $N$
one would either need an architecture with very long decoherence times, or very strong couplings.  The protocols in Secs.~\ref{1d} through \ref{2d} do not suffer from this difficulty.

\section{Other applications}\label{other}
Variants of the gates discussed here have utility beyond state preparation.  In particular, the {literature contains a number of works that study  $U_X(\varphi=\pi)$, for which the phase angle is double the one used for preparing the ground state of the Bose-Hubbard model. } This $\pi$-gate can be considered a many-particle generalization of the two-qubit iSWAP  gate \cite{Stancil2022-sf}.  Given any bit sequence $|\sigma_1\sigma_2\cdots\sigma_L\rangle$, the $U_X(\pi)$ gate swaps the order, and applies a phase shift,
\begin{equation}
U_X(\pi)|\sigma_1\sigma_2\cdots\sigma_L\rangle=e^{i\Phi
}|\sigma_L\cdots\sigma_2\sigma_1\rangle.
\label{pi}
\end{equation}
Here $\Phi=N(L-N)\pi/2$, where $N=\sum_k\sigma_j$ is the total number of excitations.
This procedure can be useful for rapidly moving quantum information around in a large circuit.  We refer to this gate as $\Phi$SWAP$_{12\cdots n}$ where the subscript lists the linear chain of sites which are reversed by the gate.  
{
Christandl et al. considered this gate in the single particle regime
\cite{Christandle2004}, describing Eq.~(\ref{pi}) as `perfect state transfer'.  They  discussed applications to quantum information processing.  Also in the single particle regime,
 Zhang et al.  experimentally  implemented 
it
\cite{Zhang2023}.
Clark et al. noted that if  the qubits are prepared in superpositions of occupied and unoccupied states, the $\Phi$SWAP gate is entangling.  They showed how to use it, and generalizations, to produce graph states \cite{Clark2005b}.
A number of other generalizations were proposed \cite{Coutinho2019,KAY2010,Kay2017,Nguyen2010,PhysRevA.99.052115,Bose2007,Post2014,PhysRevA.85.062306,Genest2013,Genest2015}.
}

A less explored feature of these gates is that  one can chain them together produce gates which act on two sites which are separated by large distances.  For example the composition FSWAP$_{15}=\Phi$SWAP$_{2,3,4}\otimes$$\Phi$SWAP$_{1,2,3,4,5}$ will swap qubits $1$ and $5$, producing a phase which depends on the occupation of the sites between them. 
{
One interesting interpretation is that the FSWAP$_{ij}$ 
gate is related to a fermionic swap operator -- which exchanges $i$ and $j$ while multiplying by $(-1)^{\bar N}$, where $\bar N$ is the number of excitations between sites $i$ and $j$.  
Indeed, FSWAP$_{ij}$ exchanges those sites and multiplies by a number dependent phase factor $\Phi_{\bar N}$.  
If the number of particles on sites $i$ and $j$ are $n_i+n_j=0,1,2$ we find $\Phi_{\bar N}=0, \pi \bar N+(L-1)\pi/2, \pi L$, where $L=i-j+1$ is the total number of sites involved in the operation.  
Thus by combining the FSWAP with single-site phase gates on sites $i$ and $j$ one can implement the fermionic exchange.}

We argue that the multi-site $\Phi$SWAP gate can be used to quantify the role of doubly occupied sites.  
Our logic is that in the hard-core limit, time evolution under Eq.~(\ref{X}) will lead to multi-site Rabi flops between the initial state and its reflection.  Dephasing of  these Rabi oscillations can be a sensitive probe of the existence of doublons.  Note that other imperfections, including disorder and noise, will also lead to dephasing, so those effects must be controlled for.  

In principle doublons can also be detected by simply reading out each of the qubits in a circuit.  Such qubit measurements, however, typically require turning off the couplings $g$.
As already explained, most experiments are performed in the large $\eta/g$ limit, and the doublons manifold is off resonant.  If one turns off $g$ on a timescale which is slow compared to $1/\eta$, the doublons are adiabatically annihilated during the process. Our indirect measurement approach does not present this same difficulty.

{  To illustrate this doublon detection method,}
in Fig.~\ref{Fig:TransmissionDynamics} we consider time evolution from an initial product state where the $N$ left-most sites are occupied by a single particle, and all other sites are empty.  We calculate the square overlap, $P(t)$, between the time-evolved state and the mirror image of the original state.  In the hard core limit we can show
$P(t)=(\sin^{2}\theta(t))^r$ with
%\begin{equation}
$r=2\left|\sum_{j=0}^L n_j (j-L/2)\right|$, 
where $n_j=0,1$ is the number of particles on site $j$ in the initial state.  
Here $\theta(t)=\int_0^t g_X(\tau)d\tau$.
Figure~\ref{Fig:TransmissionDynamics} shows that for typical parameters the fidelity of the first SWAP is quite large{, despite the we are not in the hard core limit}.  For $N>1$, subsequent oscillations show smaller amplitudes. This is particularly well illustrated by panel (a), which focuses on the first two swaps.  The loss of fidelity is more pronounced for larger particle numbers or weaker interactions{, demonstrating the role of doublons}.

As shown in Fig.~\ref{Fig:TransmissionDynamics} (b),  the envelope of the recurrences is non-monotonic.  At short times  each subsequent peak has a smaller amplitude.  Interestingly, their amplitudes follow a roughly Gaussian envelope.  Subsequently the amplitude again increases, though the recurrence times experience a number and interaction dependent phase delay.  This is effectively a many-body interferometry experiment, and the complicated structure arises from the nontrivial spectrum of the interacting system away from the hard core limit.

\begin{figure}
    \centering
    \includegraphics[width=0.7\columnwidth]{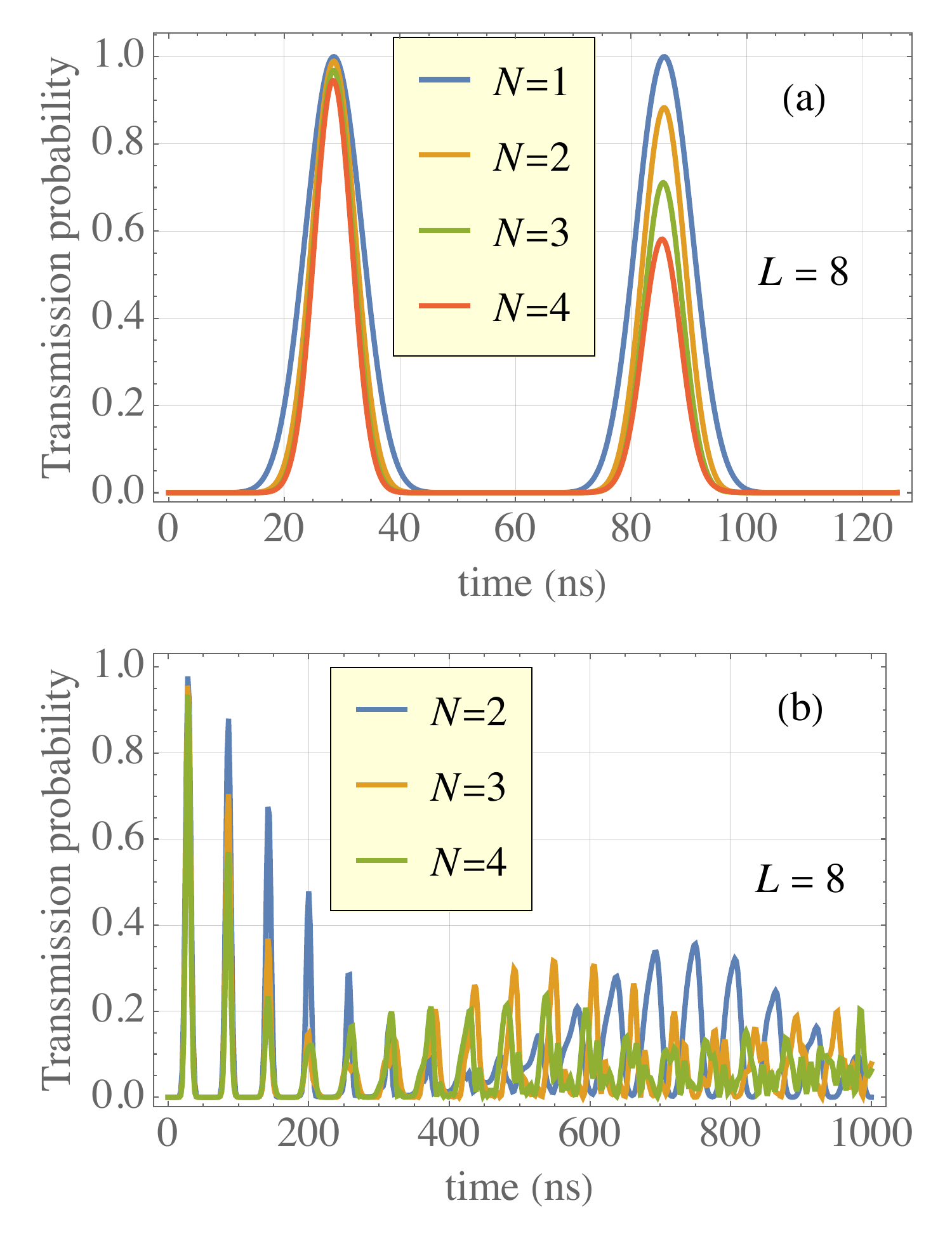}
    \caption{Transmission probability for a soft core chain with $L=8$, $g_{\rm max}/(2\pi)=35$~MHz
    and $\eta/(2\pi)=-270$~MHz at short (a) and long (b) times}
    \label{Fig:TransmissionDynamics}
\end{figure}

\section{Summary and Outlook}

Most digital quantum computation is based upon single and two-site gates.  Bespoke multi-site gates are able to perform complex tasks faster and with less control overhead \cite{katz,katz2,PRXQuantum.2.040348,cao2024multiqubitgatesschrodingercat,PhysRevA.97.042321,PhysRevA.101.022308}.  Here we introduce a set of multi-site gates which can be used for state preparation.  Constructed from $SU(2)$ and $SU(3)$ algebras, they exactly transform a localized single-particle state into an eigenstate of a hopping operator on a 1D chain or a 2D triangular lattice.
These gates also perform the exact mapping when the system can be mapped onto non-interacting particles.  In 1D such a mapping occurs either when the non-linearity $\eta$ is very small or very large -- and the gate fidelity approaches $100\%$ in both those limits.  In 2D the gate is only perfect when $\eta$ is very small.  We also constructed analogs of these gates for {\em cat states}, and discussed other applications.  For example, we observed that they can be used as part of a very sensitive many-body interferometer that probes the existence of doublons.

The key concepts used in this work was the mapping between hopping Hamiltonians and either SU(2) or SU(3) spin operators.  The SU(2) case 
is familiar from `dual rail encoding' of quantum spins -- where a spin $S/2$ is encoded in the motion of $2S+1$ bosons on two sites \cite{PhysRevA.52.3489}.  The SU(3) case is the generalization to three sites{, resulting in a 2D triangular lattice with coordination number 6}.  One could clearly extend this to higher dimensions.  { Using $SU(4)$ we can produce a model on a 3-dimensions cubic closed pack (face centered cubic) lattice with coordination number 12, forming a tetrahedron, similar to a stack of oranges in a grocery store.  The sites are labeled by integers $(n_a,n_b,n_c,n_d)$ with $n_a+n_b+n_c+n_d=\ell$. There are no obvious generalizations to lattices with other coordination numbers.}  Previous theoretical and experimental works have explored engineering hopping matrix elements to implement spin physics %\cite{Christandle2004,roy2024synthetichighangularmomentum}, 
or to manipulate quantum information \cite{Coutinho2019,KAY2010,Kay2017,Nguyen2010,PhysRevA.99.052115,Bose2007,Clark2005b,Zhang2023,Christandle2004,roy2024synthetichighangularmomentum}.

As already emphasized, state preparation is one of the most important challenges in quantum simulation.  Coherent multi-site gates can be a key part of the solution, complementing adiabatic and driven-dissipative approaches \cite{Harrington2022}.

\acknowledgments{
%PP, TK and EJM 
{
We thank Trond Andersen, Kenny Lee, Sarah Muschinske, and Will Oliver for critical comments, and important discussions about  experimental feasibility.}
We acknowledge funding from Google which enabled this work and 
support
from
the National Science Foundation under Grant No. PHY-2409403.

\appendix

\section{SU(2) Algebra}
\label{su2}
The most direct way to prove Eq.~(\ref{el1}) is to explicitly calculate the commutators between $X$ and $Z$, showing that they obey the standard $SU(2)$ algebra. Here we take a slightly more convoluted path, based upon Schwinger bosons \cite{Schwinger1952}.  This approach will generalize to the two-dimensional case in Appendix~\ref{su3}.

As introduced in the main text, we
%For this argument, we move to a first-quantized language, and consider the motion of a single particle on a 1D lattice with $L+1$ sites.  We will 
label the states of a single particle on a 1D lattice by pairs of integers which sum to $\ell$: $|0,1\rangle,|1,\ell-1\rangle,\cdots|\ell,0\rangle$.  (See Fig.~\ref{triangle}.)  We also introduce ladder operators $a,b$ which connect these state, $a|n_a,n_b\rangle=\sqrt{n_a}|n_a-1,n_b\rangle$ and $b|n_a,n_b\rangle=\sqrt{n_b}|n_a,n_b-1\rangle$.  In this one-particle subspace one can recognize that the operators in Eq~(\ref{X}) and (\ref{Z}) can be written as $X=(a^\dagger b+b^\dagger a)/2=(a^\dagger +b^\dagger)(a+b)/2-\ell$ and $Z=(a^\dagger a-b^\dagger b)/2=a^\dagger a-\ell$, where we have made use of the fact that $a^\dagger a + b^\dagger b -\ell$ vanishes for all states in our space.

We wish to use the harmonic oscillator algebra to calculate the actions of the gates $U_{X}(\theta)=\exp(-i\theta X)$ and $U_{Z}(\varphi)=\exp(-i\varphi Z)$ on the operators $X$ and $Z$.  
In that regard, we define the nested commutators by the recursion relationship 
\begin{equation}
\mathcal{L}_A^{(j)}(B)=[A,\mathcal{L}_A^{(j-1)}(B)]
\end{equation}
%{ it is better to present as a separate formula because it is used in the next section}
with the base case $\mathcal{L}^{(0)}_A(B)=B$.  Taking $X=(a^\dagger +b^\dagger)(a+b)-\ell$, we then note that for $j\geq1$, $\mathcal{L}_{X}^{(j)}(a)=(-1)^j(a+b)/2$ and hence
\begin{align}\label{aeval}
e^{-i\varphi {X}}ae^{i\varphi X}&=\sum_j \frac{(-i\varphi)^j}{j!}\mathcal{L}_{X}^{(j)}(a)\\&
=a+\frac{e^{i\varphi}-1}{2}(a+b).
\end{align}
This result is ambiguous up to an overall phase, as we could always add an arbitrary multiple of $a^\dagger a+b^\dagger b-\ell$ to $X$.  That phase drops out when we consider any physical operator.
Next we take $\varphi=\pi/2$, and note that
$e^{-i\varphi X}Ze^{i\varphi X}=e^{-i\varphi X}a^\dagger e^{i\varphi X}e^{-i\varphi X}a e^{i\varphi X}-\ell$, to find
$e^{-iX\pi/2}Ze^{iX\pi/2}=
(b^\dagger+i a^\dagger)(b-ia)-\ell$. 
Repeating the same argument with different operators yields
\begin{equation}\label{el1b}
e^{-iZ\pi/2}e^{-iX\pi/2}Ze^{iX\pi/2}e^{iZ\pi/2}=X,
\end{equation}
which as argued in the main text is a familliar result in the language of rotation operators.

\section{SU(3) Algebra}\label{su3}
We now repeat the argument from Appendix~\ref{su2} but with the two-dimensional lattice shown in Fig.~\ref{triangle}.  In particular we construct single particle operators $Q$ and $W$ with
\begin{eqnarray}\label{rtsu3}
e^{-iQ 2\pi/3}e^{-iW 2\pi/3}Qe^{iW 2\pi/3} e^{iQ 2\pi/3}&=&W.
\end{eqnarray}
The gates map the single-particle eigenstates of $Q$ onto the single particle eigenstates of $W$.

As introduced in the main text, we label each site in our triangular lattice by three integers $|n_a,n_b,n_c\rangle$ with $n_a+n_b+n_c=\ell$.  Ladder operators $a,b,c$ act on each of these integers. We take  $Q=a^\dagger a-\ell/3$ and $W=(a^\dagger+b^\dagger +c^\dagger)(a+b+c)/3-\ell/3$.

To calculate the action of the gates, we note that $[W,a]=-(a+b+c)/3$ and $[W,(a+b+c)]=-(a+b+c)$, which implies $\mathcal{L}_W^{(j)}(a)=(-1)^j(a+b+c)/3$
for $j>0$. 
Thus we find
\begin{eqnarray}
e^{-i\phi W}ae^{i\phi W}
&=&a+\frac{e^{i\phi}-1}{3}(a+b+c)\\\nonumber
&\xrightarrow{\phi=2\pi/3}&\frac{e^{2\pi i/3}-1}{3}(a e^{-2\pi i/3}+b+c).
\end{eqnarray}
The $Q$ gate then shifts the phase of the $a$ operator, giving the desired result, Eq.~(\ref{rtsu3}).

\begin{figure}[tbp]
\centering
    \includegraphics[width=\columnwidth]{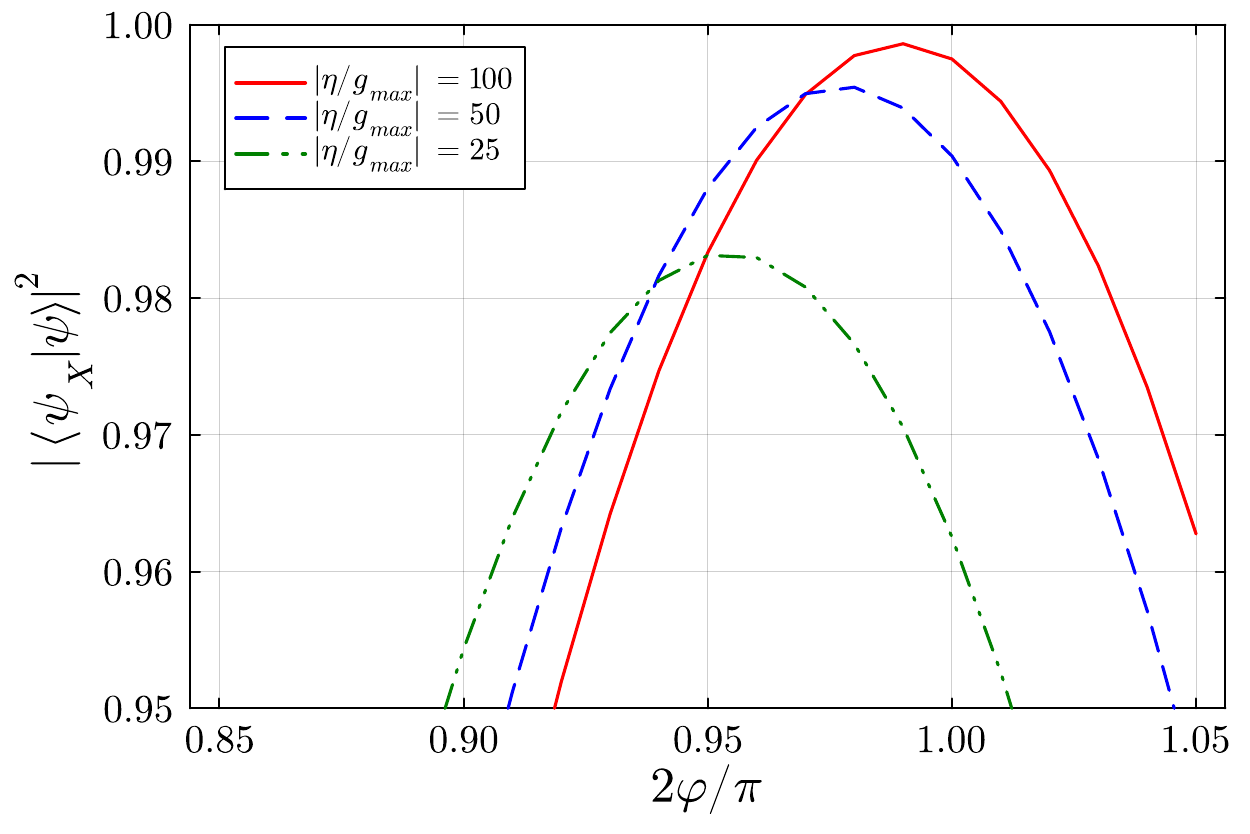}
\caption{Fidelity, $|\langle \psi_X|\psi\rangle|^2$ for the pulse sequence described in Sec.~\ref{1d}.  Here $|\psi\rangle= U_Z U_X|\psi_h\rangle$, where $|\psi_h\rangle$ is the state where $N=4$ particles occupy the left half of the chain of length $L=8$.  Here $|\psi_X\rangle$ is the ground state of Eq.~(\ref{hx}).   
The horizontal axis denotes $\varphi$, the phase accumulation in the $\tilde U_Z$.  
}\label{bhc}
\end{figure}
\section{Optimizing the phase gate}\label{optimize}

In Sec.~\ref{1d} we argued that our gate sequence produces the exact eigenstate of $H_X$ in both the hard core and non-interacting limit.  As shown in Fig.~\ref{FidelitySU2}, the fidelity is also large for a wide range of parameters away from these ideal limits.  It is natural to ask if the fidelity can be improved by modifying the gates.  The simplest such modification is adjusting the phase angles $\theta$ and $\varphi$.

In Fig.~\ref{bhc}
we show the gate fidelity as a function of 
$\varphi$, fixing $\theta=\pi/2$. 
When $\eta/g\to\infty$ the optimal phase angle is $\varphi^*=\pi/2$. 
As one decreases $\eta/gg$ the optimal angle shifts:
Fig.~\ref{bhc} shows that $\varphi^*$ falls as one moves away from the hard core limit.
We understand this result by noting that during time evolution one encounters a spatially dependent density profile (see Fig.~(\ref{density}).    For early times the density is higher on the left part of the system, and interactions produce a phase gradient, which must be corrected by adjusting $\varphi$. 
We find almost no advantage to shifting $\theta$ away from $\pi/2$.  This approach can be viewed as a highly specialized variational quantum eigensolver \cite{Tilly2022}{, or an application of the Quantum Approximate Optimization Algorithm \cite{farhi2014quantumapproximateoptimizationalgorithm}.}

The $\eta$ dependence of $\phi^*$ is also illustrated in Fig.~\ref{Fig:COM}, where we analyze a protocol for monitoring the success of our state preparation protocol.  One can experimentally optimize $\phi$ by monitoring how the Rabi oscillations are influenced by the gate time.

\section{Relationship between eigenstates of $H_X$ and eigenstates of uniform hopping Hamiltonian}
\label{sec:uniform}

The central result of our paper is that we can use a multi-site gate to produce the ground state of $H_X$ (or $H_W$ in 2D).  This Hamiltonian is a Bose-Hubbard model with inhomogeneous hopping matrix elements.  An obvious concern is that this goal could be perceived as  somewhat unnatural.  Surely it would be preferable to produce the ground state of Eq.~(\ref{bhm}) with uniform hoppings.  Those concerns are alleviated by noting that, in the strongly interacting limit, the many-body ground state of Eq.~(\ref{hx}), $|\psi_X\rangle$, has an extremely large overlap with the ground state of the {\em uniform} hopping model $|\psi_{\rm u}\rangle$.  In Fig.~\ref{fid} we show the overlap $|\langle \psi_X|\psi_{\rm u}\rangle|^2$ for $N=L/2$ hard core particles in a chain of length $L$.  The overlap is large, even for thousands of sites.  One can further alleviate any concerns by following up our gate sequence with an adiabatic evolution step, where the couplings $g_{ij}$ are slowly tuned to a uniform value. { The conditions for  adiabaticity are less stringent here, compared with transforming an insulator into a superfluid, as the initial and final state are closer together.} 

We emphasize that even though the $|\psi_u\rangle$ is almost identical to $|\psi_X\rangle$, the non-uniform hopping is an essential part of the gate. One cannot simply replace $g_{ij}$ with a constant.

To calculate the fidelity in Fig.~\ref{fid}, we use a Jordan-Wigner transformation to write the wavefunctions as the absolute value of Slater determinants.  We then note that $\langle \psi_X|\psi_{\rm u}\rangle$ is equal to the determinant of the overlaps between the single particle states which make up those  %Slater 
determinants \cite{PhysRev.97.1474}.
%made from the $N$ lowest energy single particle sites.
We write the single-particle Hamiltonian as a $L\times L$ matrix, numerically finding the eigenstates, the overlaps, and the determinant of the $N\times N$ matrix that is constructed from them.  

\begin{figure}
 \centering
 \includegraphics[width=\columnwidth]{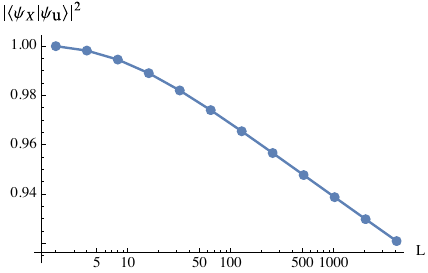}
 \caption{
 Overlap between the many-body ground state of the Hamiltonian $H_X$ and the uniform hopping model for $N$ hard-core bosons on a 1D chain of length $L$.  Here $N=L/2$.
 % Fidelity, $|\langle \psi_{\rm uniform}|\psi\rangle|^2$ for the pulse sequence described in Sec.~\ref{1d}, in the hard-core limit.  Here $|\psi\rangle=U_Z(\pi/2)U_X(\pi/2)|\psi_0\rangle$, where $|\psi_0\rangle$ is the state where particles occupy the left half of the system.  The ground state of the model with uniform hopping is $|\psi_{\rm uniform}\rangle$.  Here $L$ is the number of sites, and $N=L/2$ is the number of particles.
 }\label{fid}
 \end{figure}

\section{Numerical techniques}\label{numerics}
For most of our numerical results we use matrix product state ansatz for $|\psi\rangle$. Ground states are found by using the Density Matrix Renormalization Group algorithm \cite{Schollwck2011}.  For time evolution we use a variant of the Time Evolving Block Decimation algorithm \cite{vidal}, subdividing each gate into 200 Trotter steps, during which we implement a sequence of 2-site gates.  We systematically increased our bond-dimension cutoff and our number of Trotter steps until we achieved convergence.  %{  Add bond dimensions?} 
We use the ITensor library to implement these  algorithms \cite{itensor}. 

For small system sizes we instead use ``exact diagonalization''  where we simply enumerate the whole physics Hilbert space and write operators as finite matrices.  For example, if we have $1$ particle in 8 sites, there are only $8$ states in the Hilbert space and this approach is efficient.  We then used a split-step algorithm for time evolution.

\bibliography{bhm.bib}
%include{bhm.bbl}

\end{document}